# Binary Particle Swarm Optimization based Biclustering of Web usage Data

R.Rathipriya
Department of Computer Science , Periyar University, Salem, Tamilnadu, India

K.Thangavel
Department of Computer Science , Periyar University, Salem, Tamilnadu, India

J.Bagyamani
Department of Computer Science , Government Arts College, Dharmapuri, Tamilnadu, India

## ABSTRACT
Web mining is the nontrivial process to discover valid, novel, potentially useful knowledge from web data using the data mining techniques or methods. It may give information that is useful for improving the services offered by web portals and information access and retrieval tools. With the rapid development of biclustering, more researchers have applied the biclustering technique to different fields in recent years. When biclustering approach is applied to the web usage data it automatically captures the hidden browsing patterns from it in the form of biclusters. In this work, swarm intelligent technique is combined with biclustering approach to propose an algorithm called Binary Particle Swarm Optimization (BPSO) based Biclustering for Web Usage Data. The main objective of this algorithm is to retrieve the global optimal bicluster from the web usage data. These biclusters contain relationships between web users and web pages which are useful for the E-Commerce applications like web advertising and marketing. Experiments are conducted on real dataset to prove the efficiency of the proposed algorithms.

## 1. INTRODUCTION
Web usage mining is an important research area of web mining. It is used to analyze the user patterns from web usage data [1]. Usually, web usage data is collected from web server which is in the form log file. The mined knowledge from web log file is used by the companies to establish better customer relationship by giving them exactly what they need. The companies can find, attract and retain customers which help them to improve their business performances.

In marketing and advertising, a target audience is a group of customers that the business has decided to aim its marketing efforts and ultimately its merchandise. Target markets are groups of customers separated by distinguishable and noticeable aspects. Target Marketing is the most effective way to communicate with the customers [2]. A widely used data mining technique for analyzing the web data to create target markets is clustering. Clustering of web usage into user groups or page groups help the companies to know their browsing patterns by which they target the users.

The main characteristic of the web usage data is that it is not necessary to include all the users or pages in the clusters. Infact, it may be more useful to identify a subset of pages where a subset of the users acts in a coherent manner. Generally, clustering[3] is used to group users who have similar browsing interest under multiple pages of a web site or to group pages based on the similar browsing interest of the users. It is based on the assumption that all related users behave similarly across all set of pages of a website and vice versa. In the most cases, web users behave similarly only on a subset of pages and their behavior is uncorrelated over the rest of the conditions. Therefore, traditional clustering methods will fail to identify such users groups.

Consider the user access data matrix shown in Figure 1. If we consider all pages, users 1, 2, and 4 do not seem to behave similarly since their hit count values are uncorrelated under page 2 ,while users 1 and 2 have an increased hit count value from page 1 to page 2, the hits of user 4 drops from page 1 to page 2. However, these users behave similarly under pages 1, 3, and 4 since all their hit count values increase from page 1 to page 3 and increase again for page 4. A traditional clustering method will fail to recognize such a cluster since the method requires the three users to behave similarly under all pages which are not the case.

| | Page1 | Page 2 | Page 3 | Page. 4 |
|---|---|---|---|---|
| User 1 | **0.0** | 5.0 | **3.0** | **6.0** |
| User 2 | **1.0** | 20.0 | **4.0** | **7.0** |
| User 3 | 10.0 | 10.0 | 20.0 | 6.0 |
| User 4 | **5.0** | 0.0 | **8.0** | **11.0** |

Figure-1. A sample user access data matrix and a hidden bicluster.

To address the limitation of traditional clustering, the biclustering method [4] was introduced. In contrast to traditional clustering, a biclustering method produces biclusters, each of which identifies a set of users and a set of pages under which these users behave similarly. For example, an appropriate biclustering method will recognize highlighted hidden bicluster from Figure-1. These biclusters reveal the local browsing patterns of related users along related set of pages of a website.

The main focus of this paper is to introduce a Binary Particle Swarm Optimization (BPSO) based biclustering algorithm for web usage data. The objective of this algorithm is to identify the maximal subset of users which shows similar browsing pattern under a specific subset of pages. The information provided by these biclusters can be used in Target Marketing, Recommendation Systems and Electoral Data Analysis.

Recommendation Systems and Target Marketing are the important applications in E-Commerce [4]. In these applications, the main goal is to identify group of web users or customers with similar behavior so that one can predict the customer's interest and make proper recommendations to improve their performance. It provides a way to find users' behavior, their preferences and desires in order to provide each user excellent and personalized web services at low costs. It is vital to analyze the obtained results by eliminating the irrelative deduction rules or patterns and extracting the interesting ones. This analysis provides valuable information on how to improve structure of a website and to enhance the target marketing strategy based on the recommender system.

The remaining part of the paper is organized as follows. Section 2 provides a brief overview of existing work in the literature. Section 3 describes the methods and material for Binary PSO based biclustering. The proposed algorithm is described in the Section 4. Experiment carried out in this paper is described and presents the results in the Section 5. Summary of the paper and suggestion for the future work is given Section 6.

## 2. RELATED RESEARCH

More than one decade, many works [3,5] have been suggested to analyze the voluminous web usage data by using web usage mining techniques. Cooley et al [1], introduced methods for preprocessing the user log data, web page reference model and user navigation behavior model using web server log. Web clustering approach based on a distance function to identify the objects (either user or page) that are clustered together (similarity based) or to other probabilistic techniques called model based clustering.

In the literature, most of the paper used clustering technique for analyzing the web usage data. These results are used for personalization of web sites, system improvement, modification of web site and usage categorization. These clustering techniques are mainly operated on one dimension of the web usage data only i.e. user or page rather than taking into account the correlation between web users and pages. But in the most cases, the users are interested only in the subset of pages rather than entire set of pages. In [5], biclustering approach based on fuzzy is proposed i.e. web data elements are assigned to one or more biclusters with different membership levels and interpretation of the biclustering results may be useful for web based applications. But it fails to capture the coherent browsing behavior of the users across the subset of pages.

In the web mining area, there is no related work that has applied a specific biclustering algorithm for targeting the right group of users/ pages for marketing strategy. Madeira and Oliveira [4] have reported in their survey, that biclustering algorithms can be used for target marketing field.

For some of the applications like individualized marketing for dynamic recommendation, biclustering technique is more useful than traditional clustering technique. This is because that clustering is based on the entire users or pages but discovery of subset of pages where a subset of the users acts in a coherent manner is more useful. This type of results is provided by the biclustering techniques.

## 3. OVERVIEW OF BICLUSTERING APPROACH

### 3.1 Bicluster Types

A bicluster of a web usage data is defined as a subset of users which exhibit similar interest or browsing patterns along a subset of pages. The rows and columns of the bicluster need not be contiguous as in the user access matrix A.

The different types of biclusters[4] are Constant bicluster, Constant row, Constant column bicluster, Additive bicluster, Multiplicative bicluster, Coherent bicluster and Coherent evolution bicluster. The more meaningful type of bicluster for E-Commerce application is the coherent bicluster. This type of bicluster captures the highly correlated behaviour of the user over the subset of web pages. Therefore, the proposed algorithm is designed and implemented in MATLAB tool to discover the coherent biclusters from the web usage data.

### 3.2 Coherent Bicluster

A bicluster with coherent values is defined as the subset of users and subsets of pages with coherent values on both dimensions of the user access matrix A. In literature, Average Correlation Value (ACV)[6] is used widely to measure the degree of coherence of the biclusters.

### 3.3 Clickstream Data Pattern

Clickstream data [7] is one of the web usage data format. It is defined as a sequence of Uniform Resource Locators (URLs) browsed by the user within a particular period of time. To discover pattern of group of users with similar interest and motivation for visiting the particular website can be found by analyzing the clickstream data. It cannot use as such, it requires the some preprocessing before it is taken for analyze.

### 3.4 Preprocessing of Clickstream Data Pattern

Clickstream data pattern is converted into web user access matrix A by using equation (1) in which rows represent users and columns represent pages of web sites. Let A (U, P) be an 'n x m' user access matrix where U be a set of users , P be a set of pages of a web site, n' be the number of web user and 'm' be the number of web pages. It is used to describe the relationship between web pages and users who access these web pages. The element $a_{ij}$ of A(U,P) represents frequency of the user $U_i$ of U visit the page $P_j$ of P during a given period of time.

$$a_{ij} = \begin{cases} Hits(u_i, p_j), if\ p_{j\ is\ visited\ by\ u_i} \\ 0, otherwise \end{cases} \quad (1)$$

where Hits($U_i$ ,$P_j$) is the count/frequency of the user $U_i$ accesses the page $P_j$ during a given period of time.

## 3.5 Initial Biclusters

Given a web user access matrix A, let $k_u$ be the number of clusters on user dimension and $k_p$ be the number of clusters on page dimension after K-Means clustering is applied. $C^u$ is the set of user clusters and $C^p$ is the set of page clusters. Let $c_i^u$ be a subset of users and $c_i^u$ □ $C^u$ ($1 \leq i \leq k_u$). Let $c_j^p$ be a subset of pages and $c_j^p$ □ $C^p$ ($1 \leq j \leq k_p$). The pair ($c_i^u$ , $c_j^p$) denotes a bicluster of A . By combining the results of user dimensional clustering and page dimensional clustering, $k_u \times k_p$ biclusters are obtained. These correlated biclusters are called initial biclusters.

## 3.6 Encoding of Biclusters

Each initial bicluster is encoded as a binary string [8, 9]. The length of the string is the number of rows plus the number of columns of the user access matrix A (U, P) ie. n + m, where n and m are the number of rows (users) and of columns (pages) of the user access matrix, respectively. Each of the first 'n' bits of the binary string is related to the rows, in the order in which the bits appear in the string. In the same way, the remaining 'm' bits are related to the columns. If a bit is set to 1, it means that the relative row or column belongs to the encoded bicluster, otherwise it does not. These encoded biclusters are used as particles for binary particle swarm optimization.

| | | | | | | | | |
|---|---|---|---|---|---|---|---|---|
| $u_1$ | $u_2$ | … | $u_{n-1}$ | $u_n$ | $p_1$ | $p_2$ | … | $p_m$ |

**Figure 2: Encoded bicluster of length n+m**

## 3.7 Particle Swarm Optimization(PSO)

Particle Swarm Optimization (PSO) [10, 11, 12] method for optimization was first introduced by Kennedy and Eberhart and is inspired by the emergent motion of a flock of birds searching for food. It is a population-based optimization tool, which could be implemented and applied easily to solve various function optimization problems. It is used to explore the search space of a given problem to find the settings or parameters required to maximize a particular objective. The main strength of PSO is its fast convergence, which compares favorably with many global optimization algorithms like Genetic Algorithms (GA) [8, 9, 13, 14], Simulated Annealing (SA) and other global optimization algorithms.

Each particle knows its best value so far (pbest) and its position. This information is analogy of personal experiences of each particle. Moreover, each particle knows the best value so far in the group (gbest) among pbests. This information is analogy of knowledge of how the other particles around them have performed. Each particle tries to modify its position based on current positions, current velocities, distance between the current position and pbest and distance between the current position and gbest.

The PSO algorithm consists of three steps, which are repeated until stopping criteria is met :

Step 1. Evaluate the fitness of each particle

Step 2. Update personal best (pbest) of each particle, and global best (gbest)

Step 3. Update velocity and position of each particle

Fitness evaluation is conducted by supplying the candidate solution to the objective function or fitness function. Personal best (pbest) of each particle, global best(gbest) and positions are updated by comparing the newly evaluated fitness against the previous individual's pbest and gbest .

The velocity and position update step is responsible for the optimization ability of the PSO algorithm. The velocity of each particle is updated using the following equation:

$$v_i(t + 1) = w * v_i(t) + c_1r_1[pbest(x(t)) - x_i(t)] + c_2r_2[gbest(t) - x_i(t)] \quad (2)$$

$$x_i(t+1) = x_i(t) + v_i(t + 1) \quad (3)$$

The index of the particle is represented by i. Thus, $v_i(t)$ is the velocity of particle i at time t and $x_i(t)$ is the position of particle i at time t. The parameters w, $c_1$, and $c_2$ are user supplied coefficients. The values $r_1$ and $r_2$ are random values regenerated for each velocity update. The value pbest(x(t)) is the individual best candidate solution for particle i at time t, and g(t) is the particle's global best candidate solution at time t. w is inertia weight and $c_1$ and $c_2$ determine the relative influence of the social and cognitive components. Equation 3 is used to update the position of the particle. This process is repeated until the best solution is found or terminate conditions are satisfied.

However, the typical PSO is designed for continuous function optimization problems; it is not designed for discrete function optimization problems. Modified version of PSO called Binary Particle Swarm Optimization (BPSO)[15,16,17] that can be used to solve discrete function optimization problems.

The major difference between binary PSO and typical PSO is that the velocities and positions of the particles are defined in terms of the changes of probabilities and the particles are formed by integers in {0, 1}. Therefore, a particle flies in a search space restricted to zero and one. The speed of the particle must be constrained to the interval [0, 1]. A logistic sigmoid transformation function S($v_i$(t + 1)) is shown in (4) can be used to limit the speed of particle.

$$S(v_i(t+1)) = \frac{1}{1 + e^{\wedge}(v_i(t+1))} \quad (4)$$

The new position of the particle is obtained using (5) shown below:

$$x_i = \begin{cases} 1, \text{ if } r_3 < S(v_i(t+1)) \\ 0 \text{ otherwise} \end{cases} \quad (5)$$

where $r_3$ is a uniform random number in the range [0, 1].

### *3.8 Average Correlation Value*

Average Correlation Value(ACV) [6] is used to evaluate the correlation homogeneity of a bicluster. Matrix B = ($b_{ij}$) has the ACV which is defined by the following function,

$$ACV(B) = \max\left\{\frac{\sum_{i=1}^{n}\sum_{j=1}^{n}|row_{ij}| - n}{n^2 - n}, \frac{\sum_{k=1}^{m}\sum_{l=1}^{m}|col_{kl}| - m}{m^2 - m}\right\} \quad (6)$$

$row_{ij}$ is the correlation between row i and row j, $col_{kl}$ and is the correlation between column k and column l. A high ACV suggests high similarities among the rows or columns. ACV can tolerate translation as well as scaling. ACV can be used merit function to find non-perfect biclusters i.e., correlated/coherent biclusters.

### *3.9 Greedy Local Search Procedure*

A greedy algorithm [18] repeatedly executes a search procedure which tries to maximize the bicluster based on examining local conditions, with the hope that the outcome will lead to a desired outcome for the global problem. This approach employs simple strategies that are easy to implement and most of the time quite efficient. In this work, initial biclusters are enlarged and refined using Greedy Local Search procedure to get best biclusters.

## 4. BPSO BASED BICLUSTERING APPROACH

Biclustering is viewed as an optimization problem with the objective of finding the biclusters with high ACV and high volume. PSO is ideal for identifying the global optimal bicluster i.e. global optimal browsing pattern. Greedy method is used to select the row / column which increases the ACV of the bicluster. Greedy search procedure is used for growing seeds which makes a choice that maximizes the local gain in the hope that this choice will lead to globally good solution. So this type of greedy method may make wrong decisions and lose optimal biclusters[12]. That is greedy methods suffer from local maxima problem which can be eliminated by the optimization technique like PSO[4,16].

A biclustering algorithm along with Binary Particle Swarm Optimization (BPSO) technique for web usage data is proposed. The main objective of this algorithm is to identify the maximal subset of users with coherent browsing pattern under a specific subset of pages. These biclusters have high ACV value, where ACV value is the quality measure for correlation between users and pages.

BPSO is initialized with initial biclusters which is obtained by using K-Means on both the dimensions of the web access matrix A. This result in fast convergence of the gbest compared to random particle initialization of the BPSO and it also maintains high diversity in the population.

Each particle of BPSO explores a possible solution. It adjusts its flight according to its own and its companions flying experience. The personal best (pbest) position is the best solution found by the particle during the course of flight. This is denoted by the pbest and the optimal solution is attained by the global best(gbest). BPSO updates iteratively the velocity of each particle towards the pbest and gbest. For finding an optimal or near optimal solution to the problem, BPSO keeps updating the current generation of particles. This process is repeated until the stopping criterion is met or maximum number of iteration is reached.

The main goal of this work is to identify the maximal size bicluster constrained to some homogeneity threshold δ. These biclusters have high correlation among users and pages. The following fitness function is used to extract the high volume bicluster subject to ACV threshold δ.

F(I,J) is defined as

$$F(I,J) = \begin{cases} |I| * |J| \text{ , if } ACV(I,J) \geq \delta \\ 0, \text{ otherwise} \end{cases} \quad (7)$$

where |I| and |J| are number of rows and columns of bicluster and δ is defined as follows

$$\text{ACV Threshold } \delta = \frac{\sum_{p=1}^{p} \text{ACV (p)}}{|P|}$$

where |P| is the number of biclusters. Otherwise, several experimentation was conducted to choose specific constant value for threshold δ because the value of δ is less then size of the bicluster is very large at same its homogeneity is poor. The objective function is the maximization of F (I, J) which is subjected to δ.

**Algorithm : Binary PSO based biclustering algorithm**

Step 1. Generate initial biclusters using Two-Way K-Means clustering from user access matrix A.
Step 2. Improve the quality and quantity of the initial biclusters using Greedy Search procedure.
Step 3. initial particles = improved initial biclusters
Step 4. Initialize the velocity and position 'x' of each particle
Step 5. Set initial position as pbest and maximum of pbest is set as gbest.
Step 6. Repeat

```
        for each particle in the population
           current fitness =fitness(particle)
           if current fitness >
        fitness(pbest(particle))
              Update pbest with position of current
        particle
           end (if)
   end (for)
        Update gbest
        Update velocity using equation (2)
        Update position using equation (5)

        until stopping criteria is met
```

Step 7. Return gbest as global optimal bicluster

In the above algorithm, first two steps are used to create initial biclusters. Remaining steps are used to optimize the biclusters and identifies the global optimal bicluster. This global optimal bicluster holds global optimal user profile.

## 5. EXPERIMENTAL RESULT ANALYSIS

The experiments are conducted on the well-known benchmark msnbc clickstream dataset called which was collected from MSNBC.com[1] portal. This data set is taken from UCI repository, where the original data is preprocessed using equation (1). There are 989,818 users and only 17 distinct items, because these items are recorded at the level of URL category, not at page level, which greatly reduces the dimensionality.

### *5.1 Preprocessing of msnbc Dataset*

The representation of processed server log data is

i. The server log files have been converted into a set of sequences and one sequence for each user session.
ii. Each sequence is represented as an ordered list of discrete numbers
iii. Each number represents one of 17 categories of web pages requested by the user.

Each event in the sequence corresponds to a user's request for a page. Any page requests served via a caching mechanism are not recorded in the server logs and, hence, not present in the data.

These clickstream sequence is converted in to user access matrix A (U, P) using equation (1)

$$A(U,P) = \begin{pmatrix} 1\,0\,0\,0\,0\,0\,0\,0\,0\,0\,0\,0\,0\,0\,0\,0\,0 \\ 0\,2\,0\,0\,0\,0\,0\,0\,0\,0\,0\,0\,0\,0\,0\,0\,0 \\ 0\,4\,3\,1\,0\,0\,0\,0\,0\,0\,0\,0\,0\,0\,0\,0\,0 \end{pmatrix}$$

During the each user session, the user visited the page categories are marked by frequency of visit, and there are many zeros to reflect that user sessions did not visit that particular page category.

In msnbc dataset, the length of the record having less than 5 is considered as a short and record length greater than 15 is considered as a long. During data filtering process, short and long records are removed from the dataset because it does not contribute much for analyzing the user browsing behaviour. After data filtering process, number of records in the dataset is reduced to 3386 from 9 lakh records. The size of the dataset taken for the evaluation of the proposed algorithm is 3386 x 17.

Table 1: Charactertics of Initial Bicluster

| | Two Way K-Means | Greedy Search procedure |
|---|---|---|
| Average ACV | 0.4711 | 0.9913 |
| Average Volume | 494.9 | 1599.8 |

Table 2: Effect of inertia weight on global volume of bicluster (Gbest) in BPSO

| Particle size | Inertia Weight | Gbest Volume | Gbest ACV |
|---|---|---|---|
| 120 | Variable | 9712 | 0.9337 |
| 120 | Fixed w = 0.9 | 9373 | 0.9095 |
| 120 | Fixed w = 0.4 | 6415 | 0.9204 |

[1] http://www.ics.uci.edu/~mlearn/MLRepository.html

Table 3: Effect of inertia weight on volume of all biclusters (Pbest) in BPSO

| Particle size | Inertia Weight | Average pbest Volume | Average pbest ACV |
|---|---|---|---|
| 120 | Variable | 3409.5 | 0.9352 |
| 120 | Fixed w=0.9 | 5539.8 | 0.9253 |
| 120 | Fixed w=0.4 | 1822.9 | 0.9403 |

Table 4: Effect of inertia weight on Gbest based on Greedy Local Search and BPSO

| Particle size | Inertia Weight | Gbest Volume | Gbest ACV |
|---|---|---|---|
| 120 | Variable | 10208 | 0.9204 |
| 120 | Fixed w = 0.9 | 9432 | 0.9128 |
| 120 | Fixed w = 0.4 | 7884 | 0.9332 |

Table 5: Effect of inertia weight on Pbest based on Greedy Local Search and BPSO

| Particle size | Inertia Weight | Average pbest Volume | Average pbest ACV |
|---|---|---|---|
| 120 | Variable | 5112.9 | 0.9329 |
| 120 | Fixed w=0.9 | 5756.4 | 0.9281 |
| 120 | Fixed w=0.4 | 3238.4 | 0.9340 |

Inertia weight was introduced to improve the rate of convergence of the BPSO algorithm and determine how much the velocity at time 't' should influence the velocity at time 't+1'. It is ovious from the above tables, that large inertia weight facilitates the global search while the small inertia facilitates the local search. Moreover, it eliminates the setting of velocity to maximum. Variable inertia weight is used to set the balance between the global and local search abilities.

$$W = w_{max} - ((w_{max} - w_{min}) / iter_{max}) * iter \quad (8)$$

Where $w_{min} = 0.4$, $w_{max} = 0.9$, $iter_{max}$ is the maximum number of iteration and iter is the current iteration in progress.

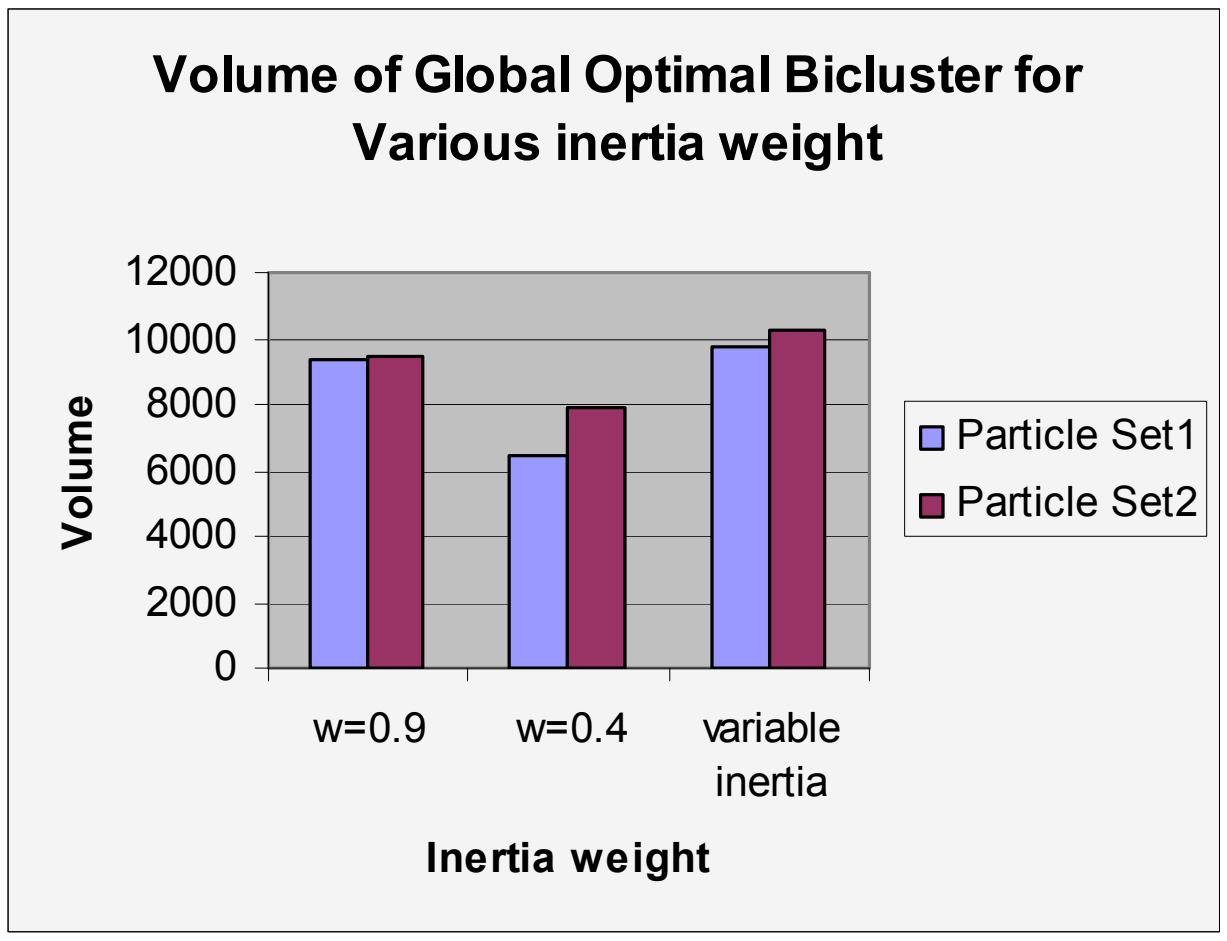

Here, initial biclusters obtained using two way K-Means algorithm is called Particle set 1. Enlarging the above initial biclusters using greedy local search procedure is called Particle set 2. It is obvious from the figure 2, optimization task with local search procedure yields good result. Binary PSO with varying inertia weight from a maximum value of 0.9 to a minimum value of 0.4 during the whole run of the algorithm. A significant of increase of volume of biclusters is achieved in variable inertia than with the fixed inertia.

The global optimal user profile consists of the following page category

| Id | Category | Id | Category |
|---|---|---|---|
| 2 | News | 10 | Living |
| 3 | Tech | 11 | Business |
| 4 | Local | 12 | Sports |
| 5 | Opinion | 14 | BBS |
| 6 | On-air | | |

A subset of users shows highly correlated browsing interest for the above subset of pages. This proposed method is useful in indentifying the target users for target marketing and thereby recommendations can be made.

Analysis of the complex relationship between users and pages of the web site is achieved through the BPSO based biclustering algorithm. The interpretation of the result is useful to answer the following questions.

- " Which are the pages that are visited together? " and
- "Which are the groups of users that exhibit coherent browsing pattern?"

It helps us to drive the marketing campaigns i.e., once coherent groups of users are identified, marketing efforts can become more focused and appropriately targeted. Thus, this biclustering algorithm is ideal for the target marketing applications.

# 6. CONCLUSION

The key contribution of this paper is to introduce biclustering algorithm with BPSO for web usage data. The objective of this algorithm is to find high volume biclusters with high degree of coherence between the users and pages. The results show that objective is attained well. The interpretation of the biclustering results can be used in the focalized marketing strategy like direct marketing and target marketing. It outperforms biclustering algorithm using greedy strategy. Moreover, identified set of biclusters cover higher percentage of users and pages and also captures the global browsing patterns from web usage data.